\documentclass[12pt]{article}
\usepackage{dn_report,graphicx,amsmath,dcolumn}
\usepackage{wrapfig,times}

\newcolumntype{d}[0]{D{.}{.}{-4}}
\newcommand{\ttmdump}[1]{#1}
\begin{document}

\ttmdump{\make_cover_page{TRI-BN-12-10} {R. Baartman} {ISAC LEBT} {The
    ISAC electrostatic Low Energy Beam Transport (LEBT) system is described.}  {July, 2012}}

\section{Introduction}
In the isotope separation on-line (ISOL) technique, radioactive ions are created at rest and accelerated with a static potential to an energy of only a few tens of keV. At this energy, the most efficient way to transport the particles to the experimenter is with electrostatic bending and focusing elements. The reason for this can be understood from the Lorentz force $\vec{F}=q(\vec{E}+\vec{v}\times\vec{B})$ ($E$ is the electric field and $B$ the magnetic; $v$ is the velocity). For $v\ll c$, and typically attainable fields, the first term is much larger than the second. Moreover, up to the sparking limit, electric fields ($\sim10$\,kV/cm) are far more economical than magnetic fields ($1$\,T). Lastly, electrostatic fields have advantages when it is required to transport ions of widely different mass and charge. Given only electric fields, all particles that have been accelerated from rest by the same potential will follow the same trajectory irrespective of mass or charge.

One is free to choose any length scale for the optics that transports and focuses the radioactive ion beam particles: widely separated quadrupoles with large apertures can in principle transport as well, with same acceptance, as small apertures and many quads, and the overall cost is little different. However, the more focusing there is per unit length, the less effected is the beam by perturbations such as stray magnetic fields and misalignments. This favours many small quadrupoles over few large ones. 

\section{Standard Modules}
In ISAC, we chose a length scale of 1 metre as a good compromise between economics and sensitivity. By this is meant that a focusing unit, consisting of two quadrupoles, has a length of 1 metre. The Twiss $\beta$-function is thus also roughly 1 metre, and beam radius for a typical emittance of $\pi\epsilon=10\pi$mm-mrad is 3\,mm. ISAC quadrupoles have length of 50\,mm and aperture of 50\,mm, with generally 25\,mm dia.\ aperture ground planes at entrance and exit. This requires the electrode voltages to be $\pm2$\,kV for a 60\,keV particle energy, and yields acceptance in the $100$ to $200\pi\mu$m range. Bends are also electrostatic, and their electrodes are spherical. This gives equal focusing in the bend and non-bend planes. Bend radius is $25.4$\,cm in all cases. There are 10 identical $90^\circ$ bend sections, consisting of two $45^\circ$ benders sandwiching a triplet tuned to make the section achromatic.

\begin{figure}[htbp]
\begin{center}
\includegraphics[height=3.1cm]{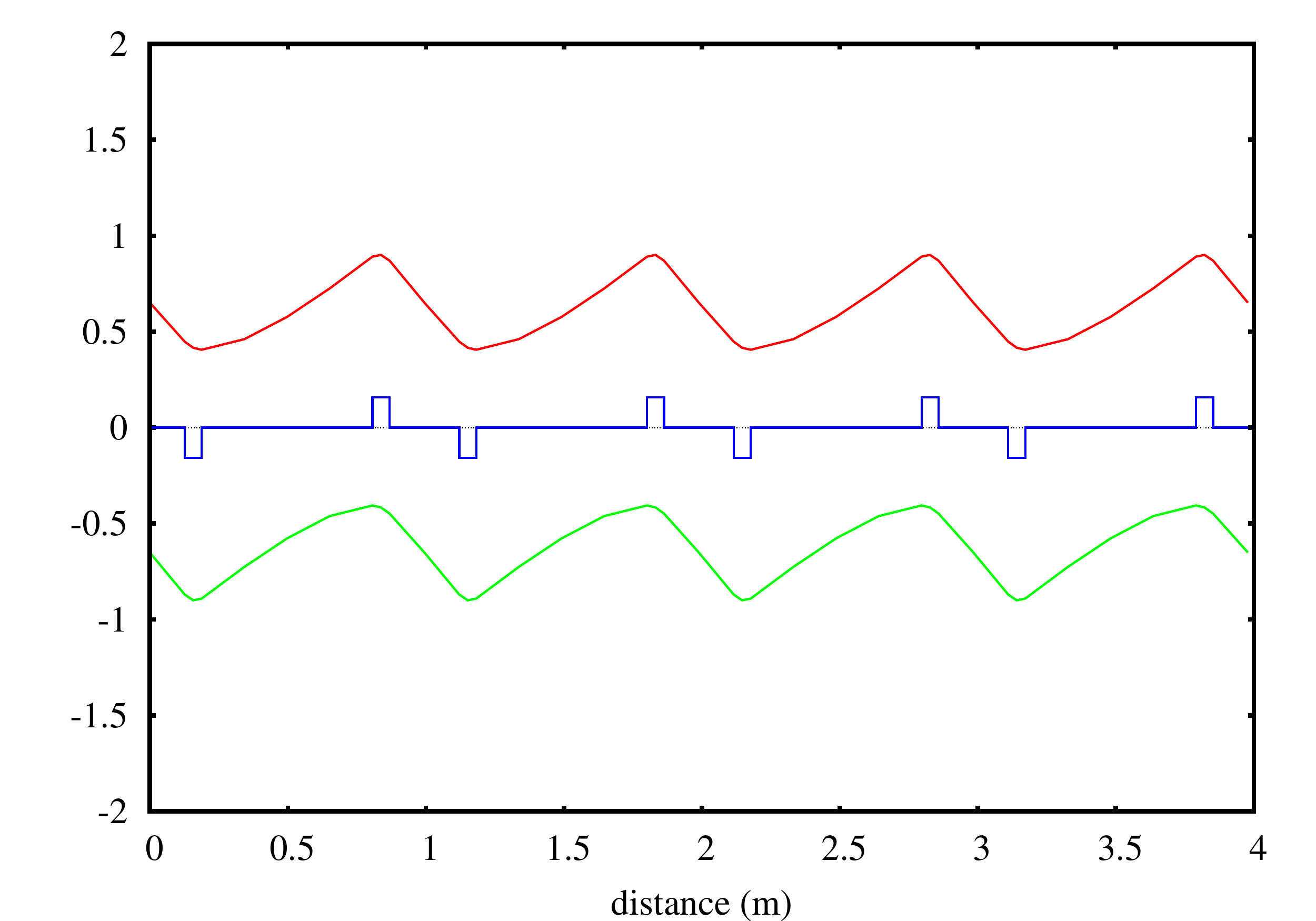}
\includegraphics[height=3.1cm]{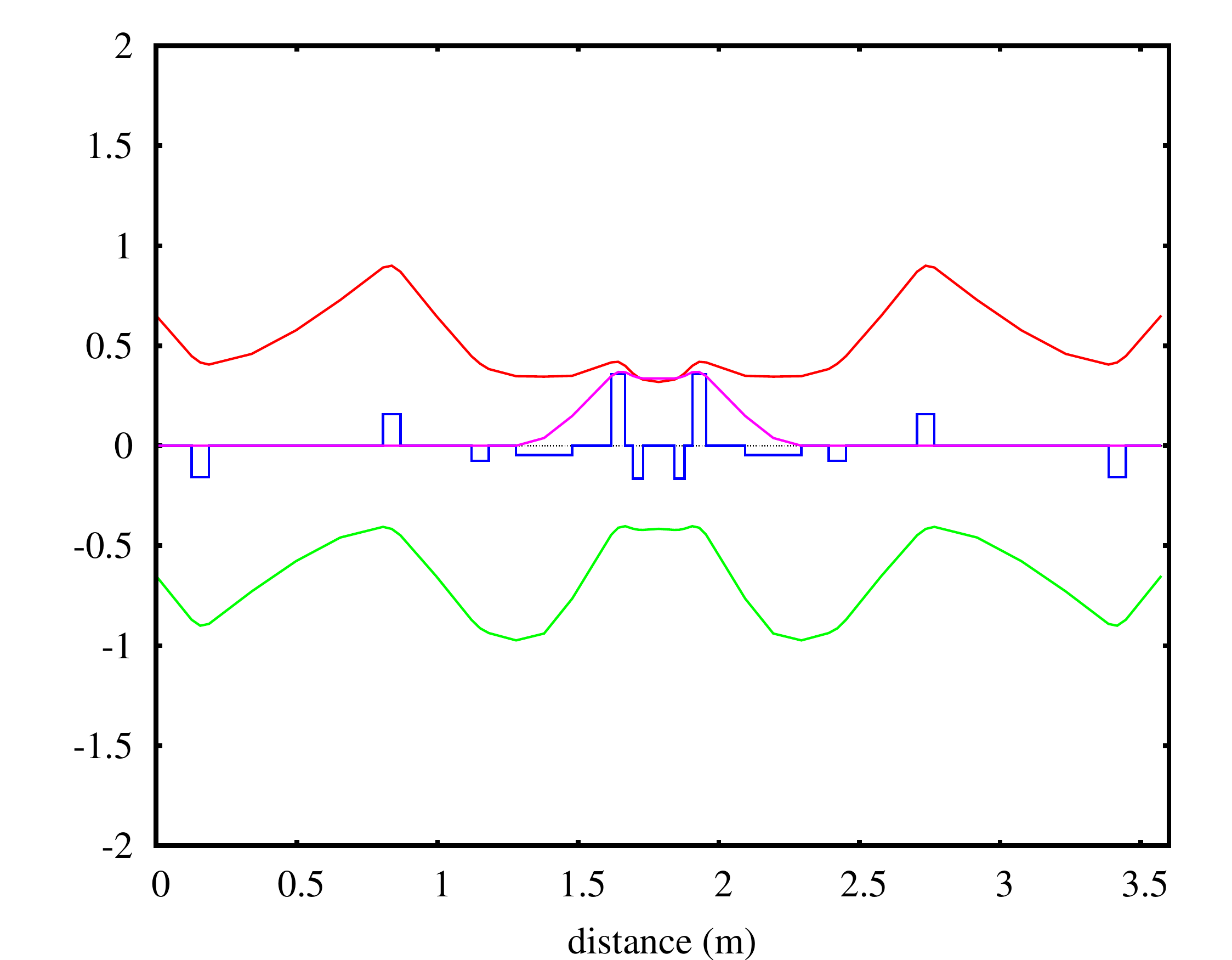}
\includegraphics[height=3.1cm]{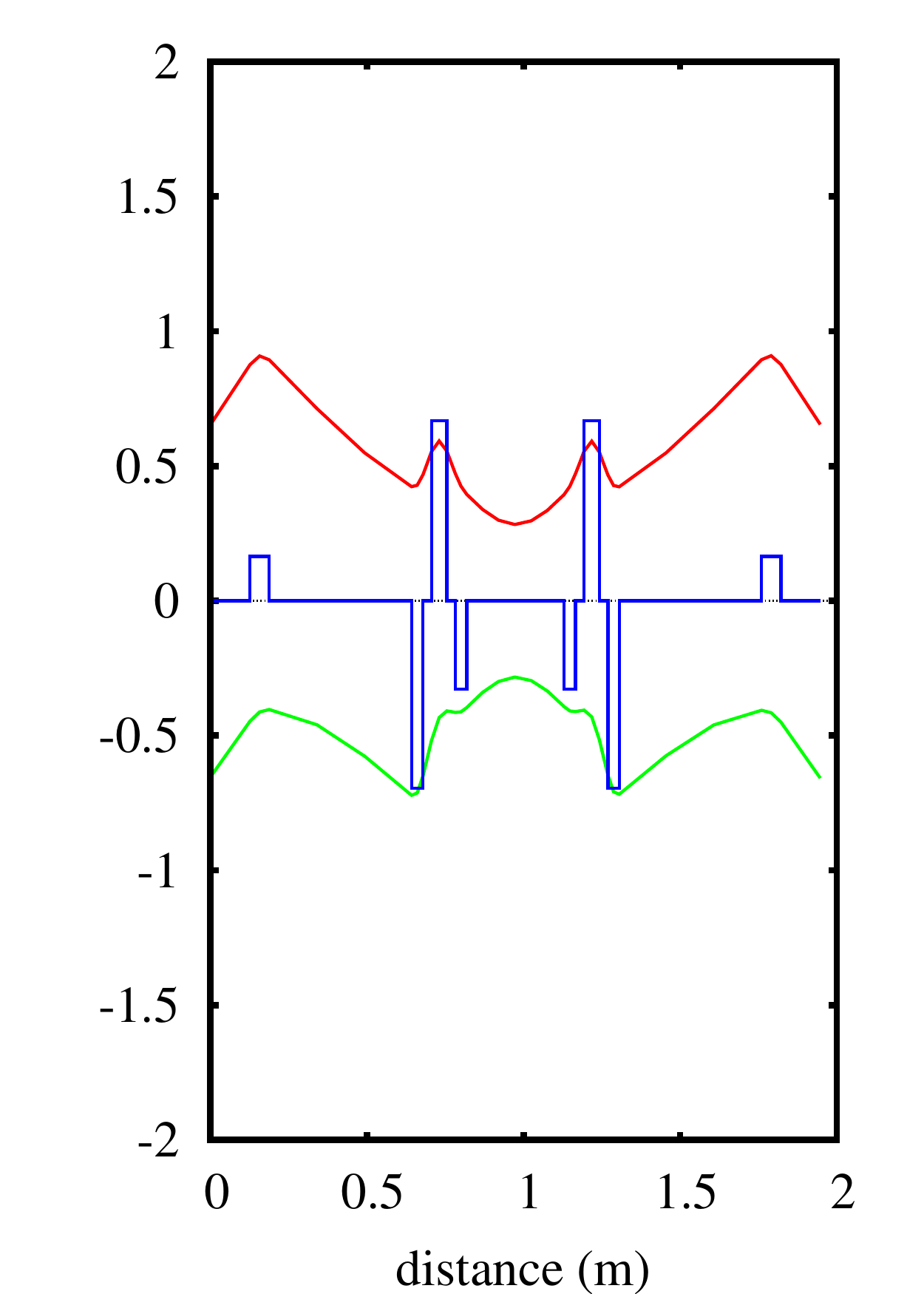}
\includegraphics[height=3.1cm]{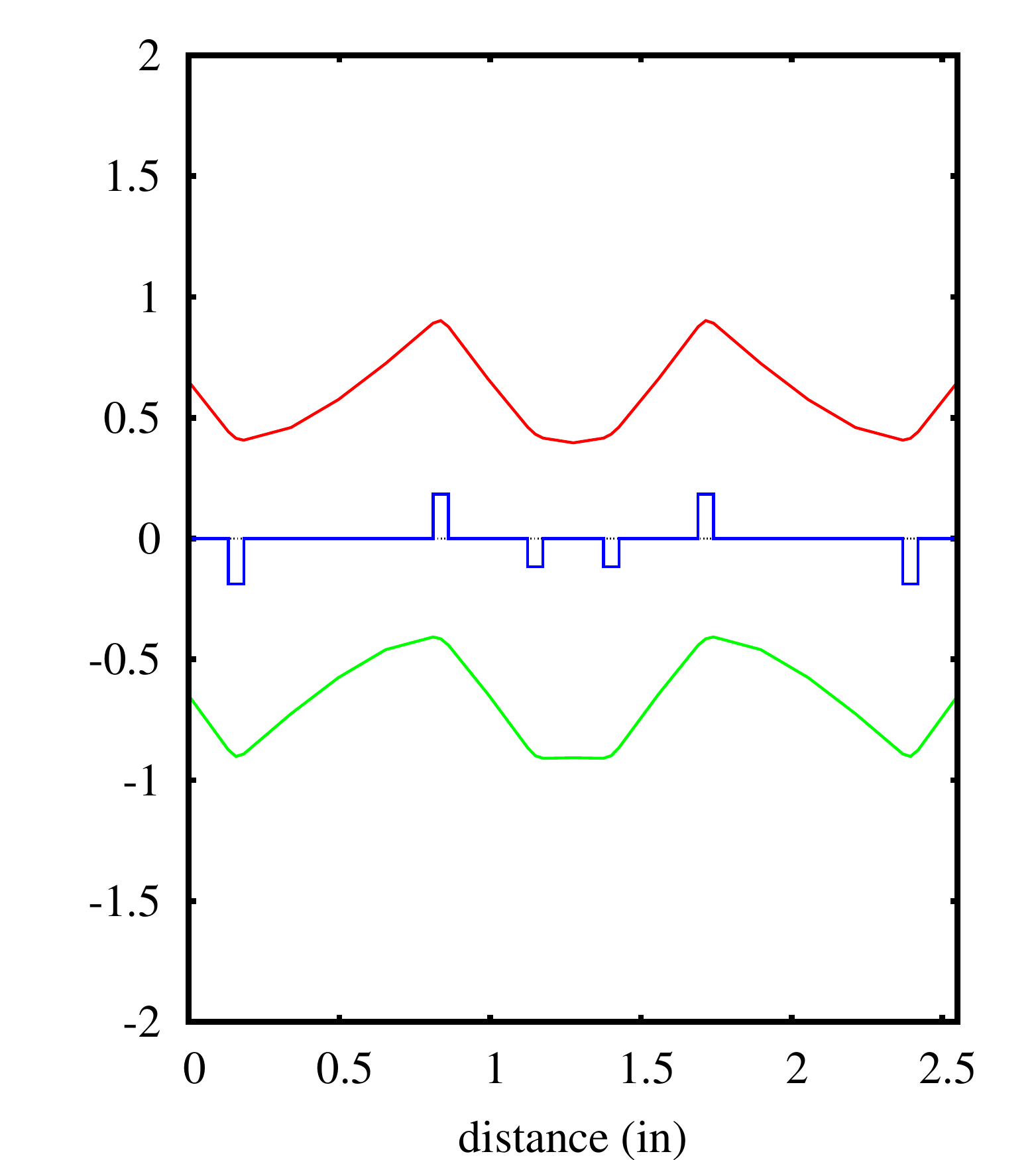}
\end{center}
\caption{Left to right, $x$ and $-y$ envelopes, in cm (red and green), for (resp.) the periodic, the $90^\circ$ bend section (dispersion in magenta), the low-$\beta$ insertion, and the order-reversing section. Emittance is $\pi\epsilon=50\,\pi\mu$m. The stepped function in blue is the focal strength in arbitrary units.}\label{env}
\end{figure}

The periodic transport sections have cells consisting of two quads; each cell has length 1.0\,m, but the quads, rather than equally separated, are arranged as doublets (see fig.\,\ref{env}). This allows that a $45^\circ$ spherical bender switch can be inserted into any long drift to create a spur line, for example to an experiment area. Insertion of the typical $90^\circ$ bend section displayed in fig.\,\ref{env} into the periodic section requires the periodic section to reverse parity (from FODO to DOFO), since the bend section is mirror-symmetric. But this same quality allows to use two $90^\circ$ bend sections in sequence if they are in opposite planes. This trick is used in bringing the ISAC beam from the separator to the main floor: from the mass separator, the beam is matched to the periodic section and needs no other matching until it arrives on the main floor, in spite of four $90^\circ$ bend sections.

In every case, the beam must be matched out of the separator, through the transport system, and matched out of the transport into the experiment. Originally, the transport system consisted of only 4 types of modules: the periodic section, the $90^\circ$ achromatic bend section, the low-$\beta$ insertion, and the order-reversing section. The low-$\beta$ section is used for the buncher of the radiofrequency accelerator (RFQ) and for the switch cross that switches between the low energy experiment area and the RFQ. The order-reversing section matches from FODO periodic section to DOFO.

\begin{figure}[htbp]\centering
\includegraphics[width=0.7\textwidth]{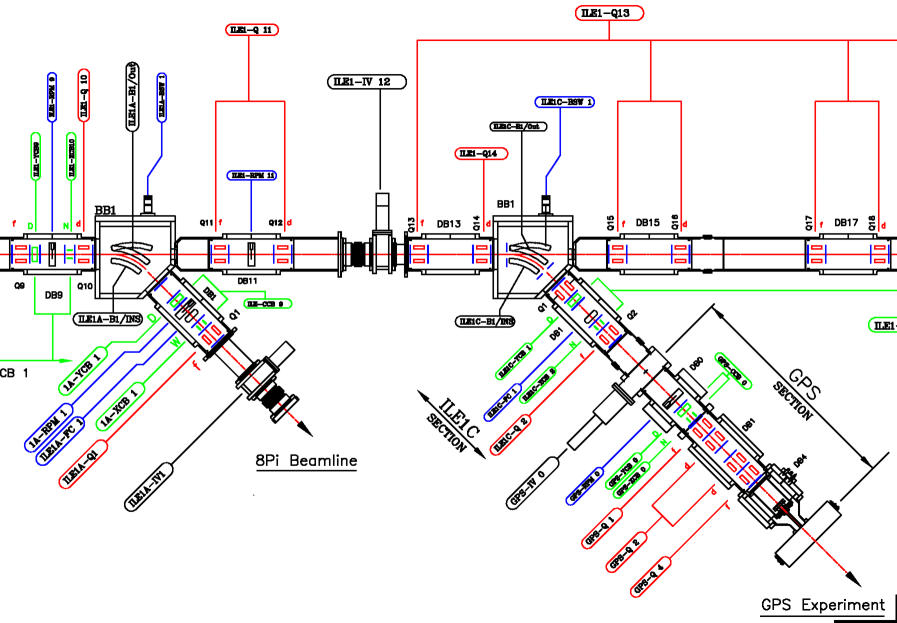}
\caption{ISAC periodic section, showing 5 periods, and 2 mechanical beam switches. Quadrupoles are red, steering correctors green, and diagnostics blue.}\label{switch}
\end{figure} 
The $45^\circ$ benders of the bend sections can function as switches, but this requires that the outer electrode move out of the way when the bend is turned off. See Fig.\,\ref{switch}; this is achieved with an air-actuated piston that pivots the electrode. This feature is used in a number of locations in ISAC: switching between an experiment such as Yield or GPS2, Francium, or carrying on straight ahead to another location.

\begin{figure}\centering
 \includegraphics[width=0.95\textwidth]{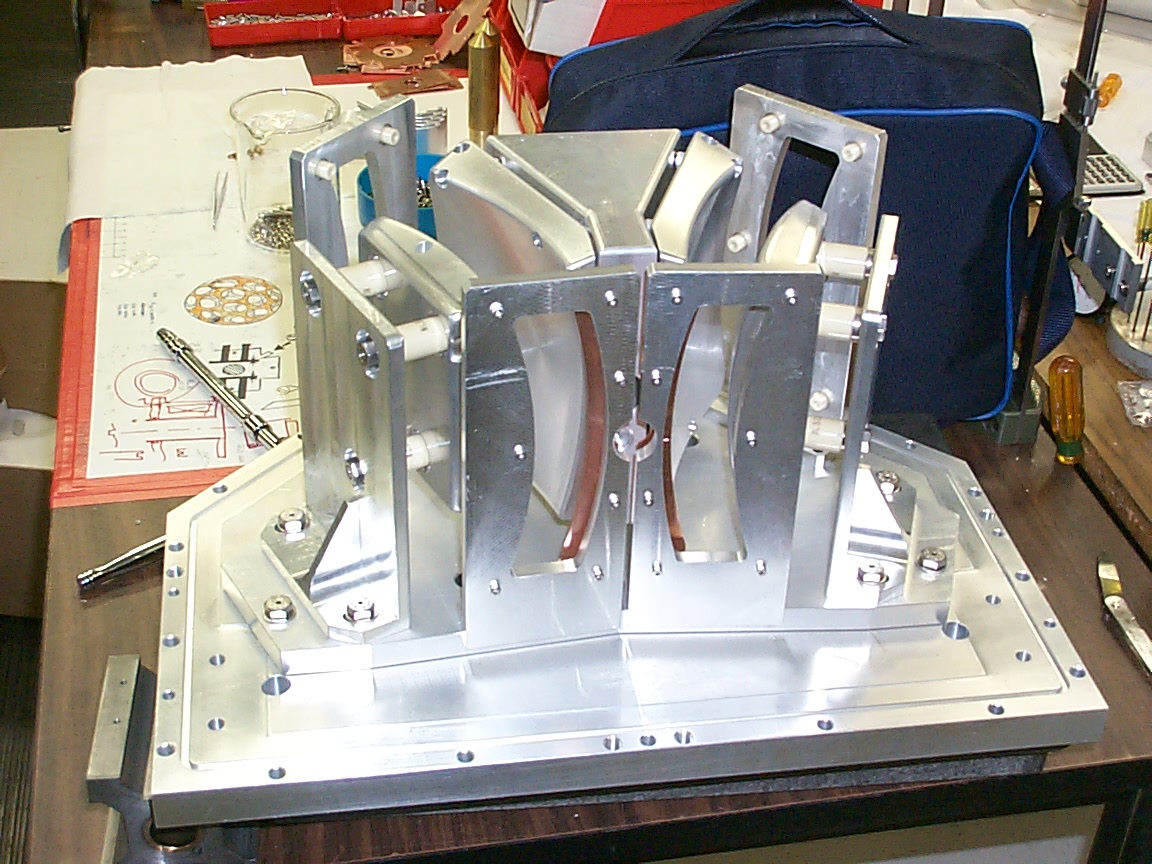}
\caption{Spherical 3-way bend unit, on the bench.}\label{three}
\end{figure}

Mechanical switching is adequate for hourly or daily switching, but too
slow for time sharing on the scale of seconds or less. Moreover,
mechanical switches do not permit co-axial overlapping of the
radioactive ion beam with for example a laser beam as is needed in the optically-pumped polarizing section. For these cases, an electrostatic switch was developed by modifying a single $45^\circ$ bender, splitting it into a $9^\circ$ electrostatic deflector and $36^\circ$ spherical bender. By combining such a $36^\circ$ bender with its mirror image, a 3-way switch was built, see fig.\,\ref{three}.
 \begin{figure}[htbp]\centering
\includegraphics[width=0.95\textwidth]{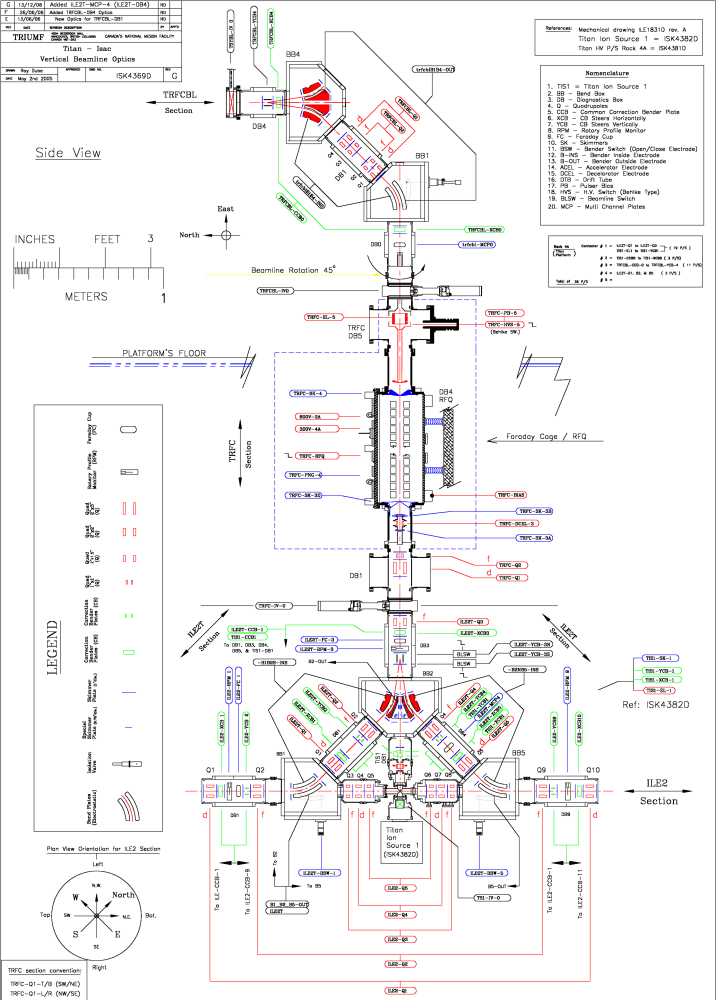}
\caption{ISAC electrostatic transport vertically to the TITAN experiment. The bottom 3-way deflector allows beam to return from the RFQ-cooler. The straight-through section sandwiched between the two 2-way mechanical benders is essentially the standard ``low-$\beta$'' insertion. The top $90^\circ$ bend section is the achromatic standard section, but distorted by the broken symmetry of having one side a $45^\circ$ bender and the other side a $36^\circ$ bender.}
\end{figure} 

\section{Special Sections}
Each experiment has its own requirements on target. Usually, these are met with a single 4-quadrupole matching section. However, there are at least four sections with special properties.

\subsection{RFQ Match}
\begin{wrapfigure}{l}{0.5\textwidth}\vspace{-1cm}
 \includegraphics[width=0.5\textwidth]{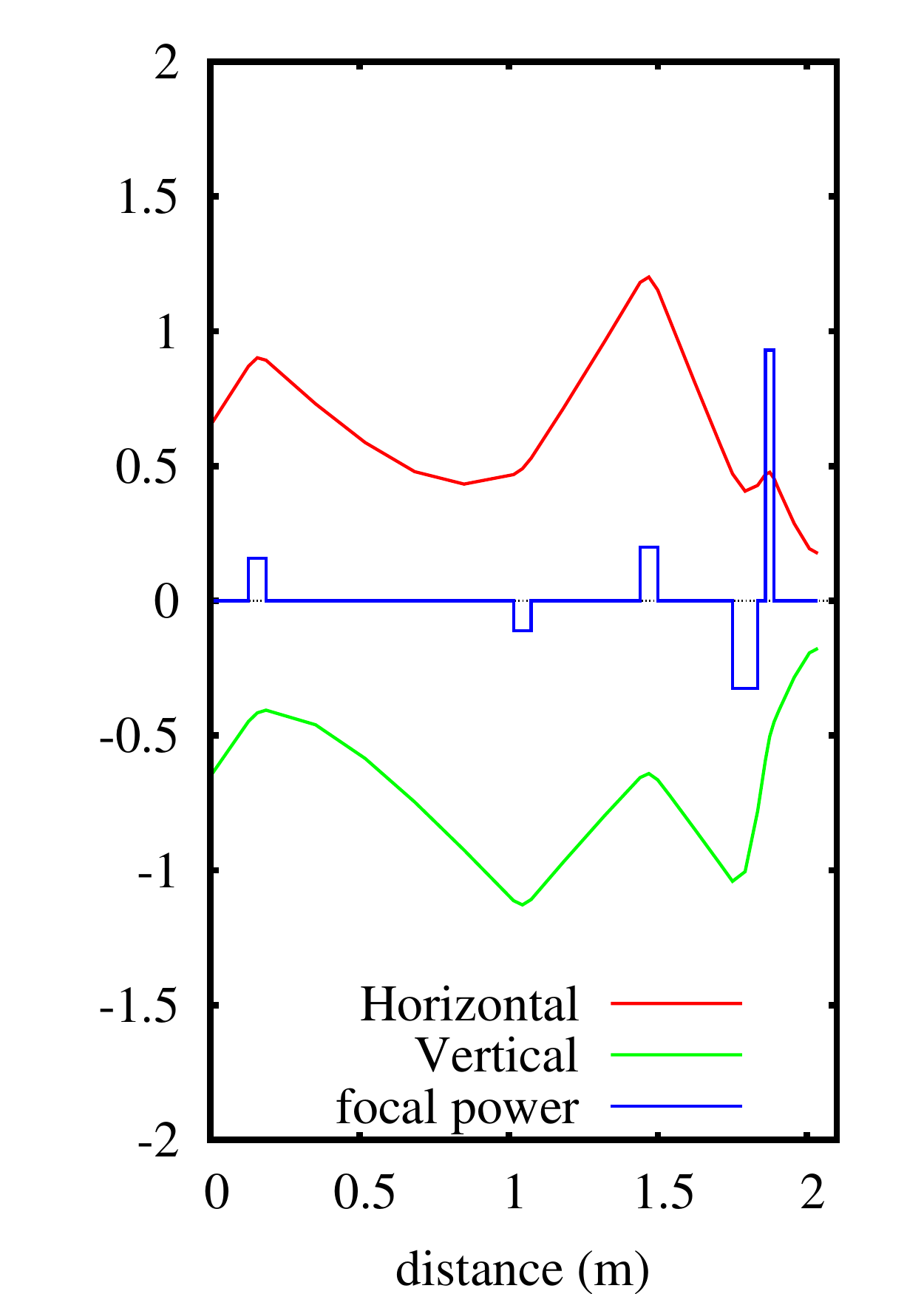}
\caption{Beam envelopes (in cm) for $\epsilon=50\,\mu$m and focal strengths (blue, in arbitrary units) of the section that matches from periodic section to the ISAC RFQ.}\vspace{-1.5cm}
\end{wrapfigure} 
Matching from the periodic section to the 150\,keV/u RFQ requires an optics scale change as the RFQ is very strongly focusing: the Twiss $\beta$-function changes from 1\,m to 6\,cm. Using only standard-sized quadrupoles, this is achievable, but the envelope in the focusing plane in the next-to-last quadrupole becomes too large and this results in unacceptably large third order aberration. The solution is to use a half-sized final quadrupole (25\,mm long by 25\,mm bore dia.), allowing the next-to-last quadrupole to be in close proximity to the match point, thus limiting the beam size and therefore the aberrations.

\subsection{$\beta$-NMR}
\begin{figure}[htbp]\centering
\includegraphics[height=3.2cm]{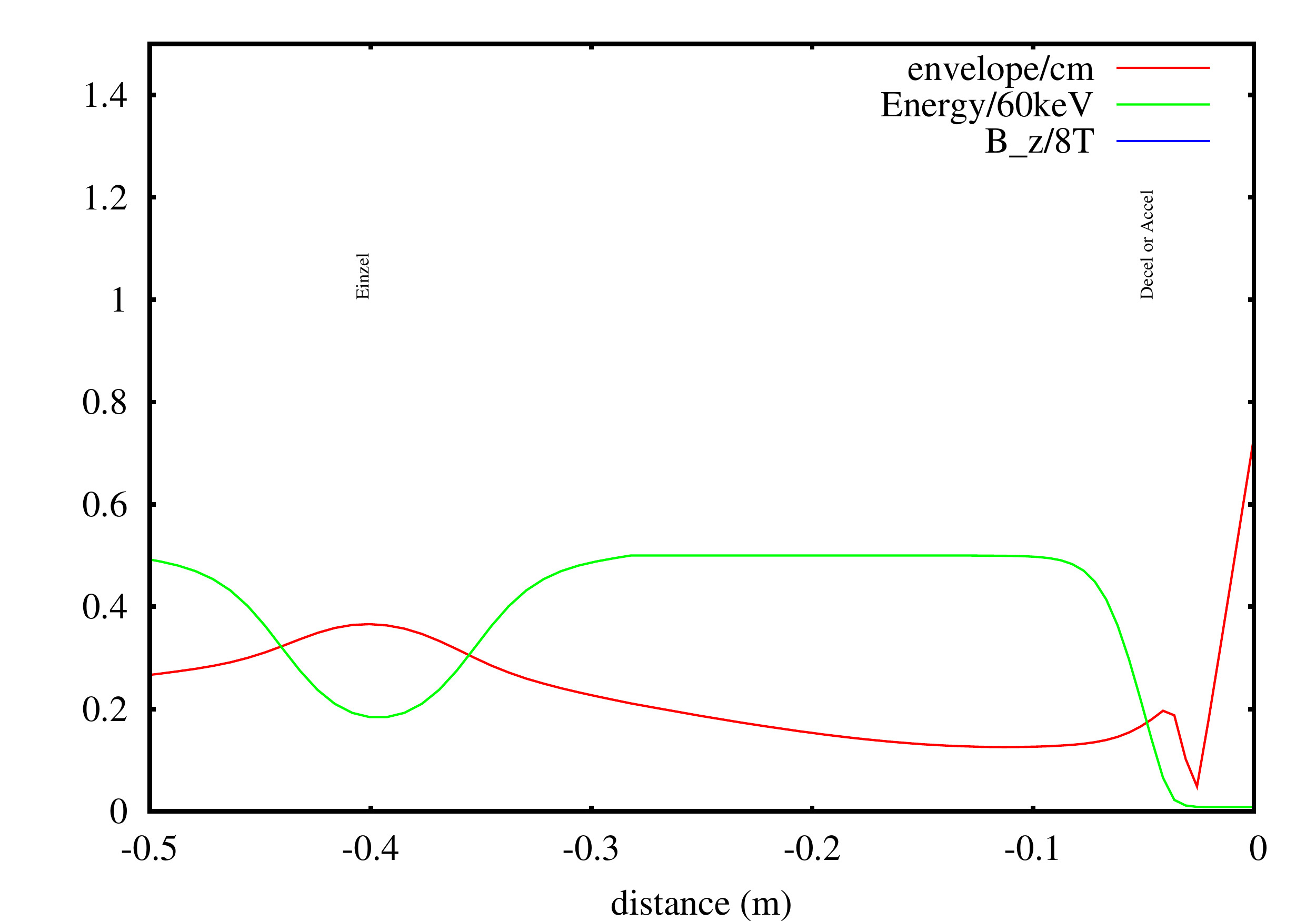}
\includegraphics[height=3.2cm]{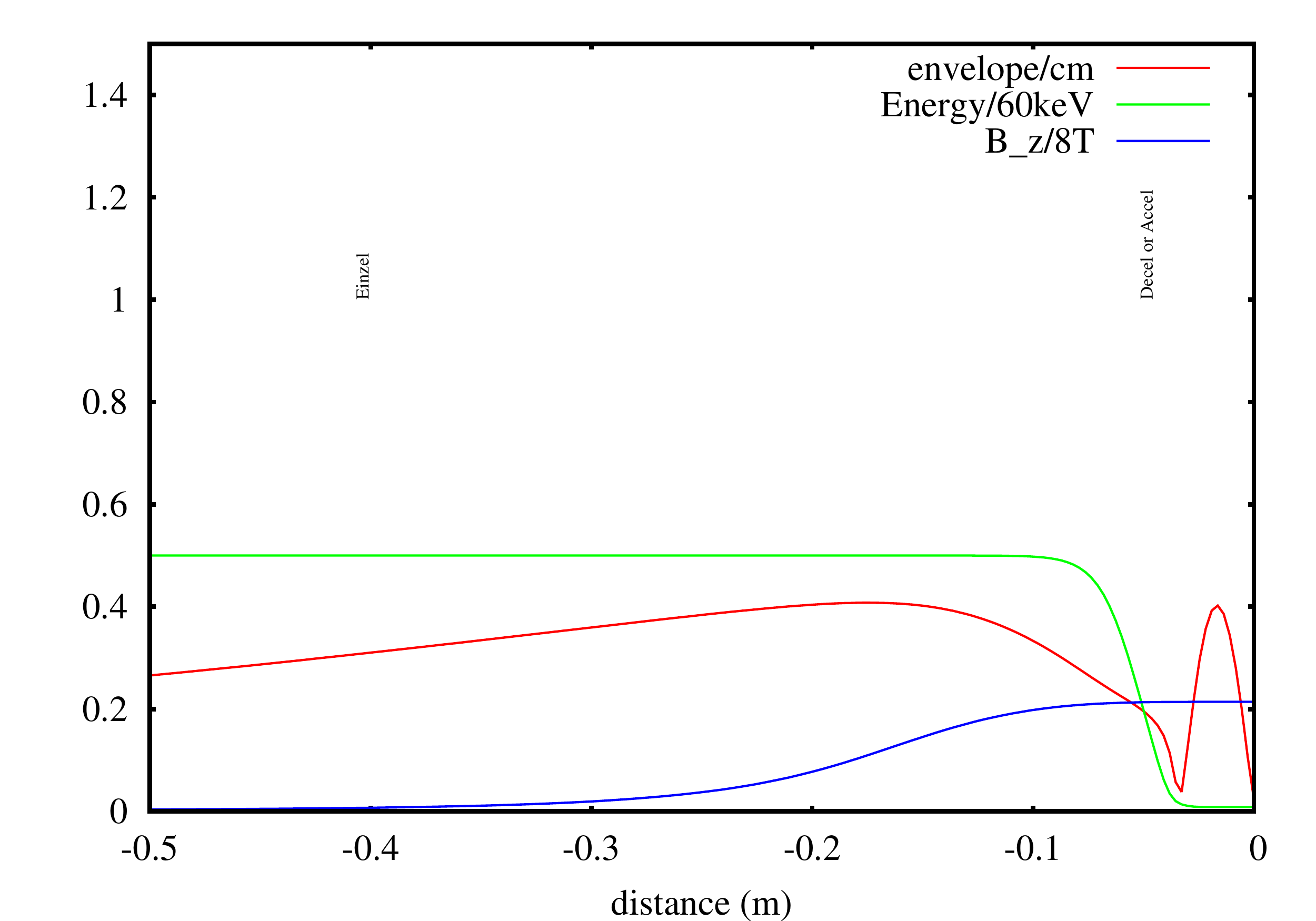}
\includegraphics[height=3.2cm]{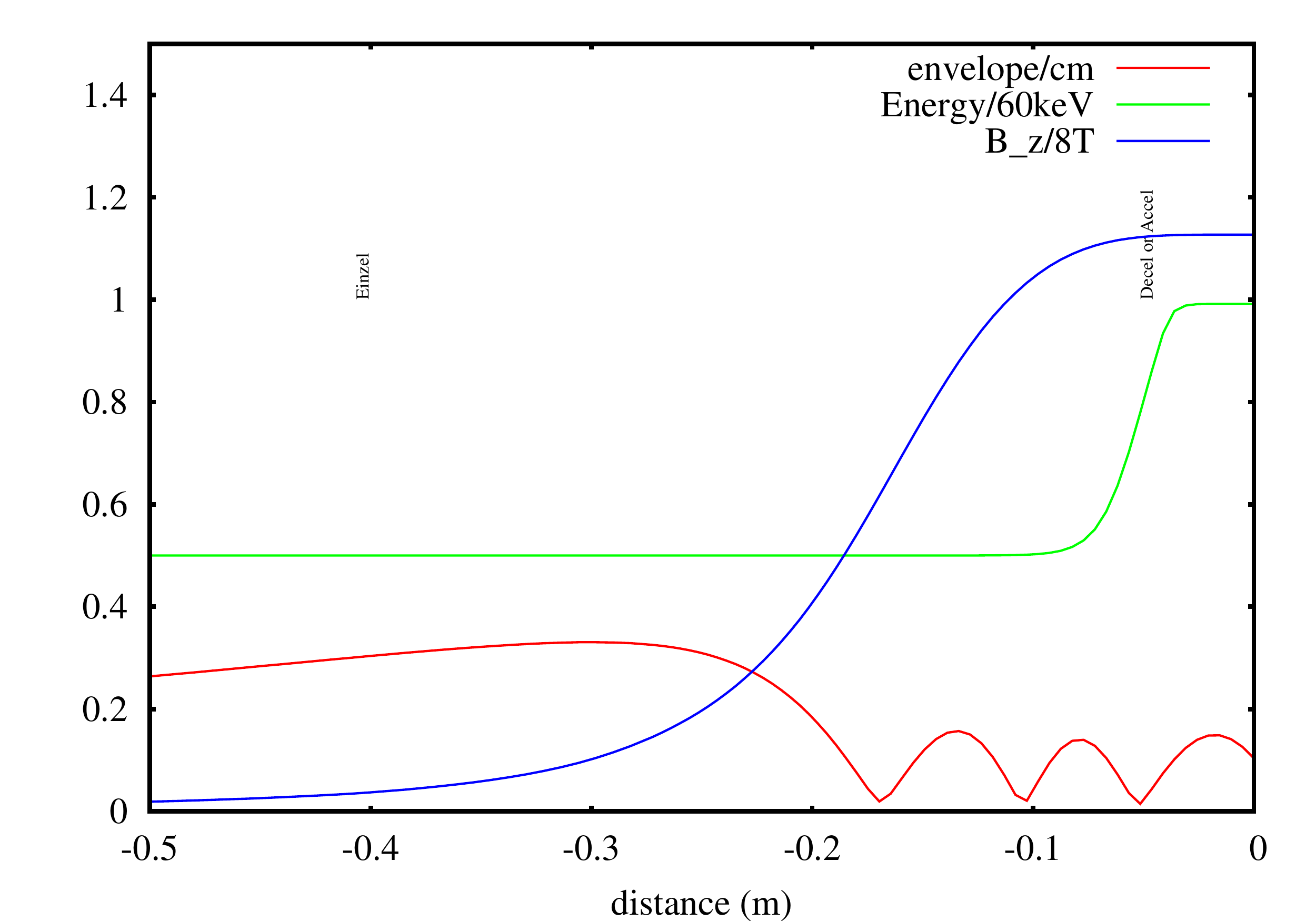}
\caption{Three other options for implanting onto the $\beta$-NMR sample. Left to right: Deceleration to 500\,eV with solenoid off, deceleration to 500\,eV with solenoid at 1.7\,Tesla, acceleration from 30\,keV to 60\,keV with solenoid at 9\,Tesla. Note: the final einzel lens at $z=-0.4$\,m cannot be used when the solenoid is powered, because it initiates a Penning discharge.}\label{f6}
\end{figure}

Requirements for the $\beta$-NMR facility are particularly challenging. Radioactive isotopes (usually $^8$Li) are to soft-land at energies down to 100\,eV, or up to 90\,keV. This is to occur in a magnetic field of up to 9\,Tesla. Einzel lenses are used for matching, as the solenoid and the deceleration electrodes are also axially symmetric. Deceleration to the lowest energies results in very strong radial electric forces which can over-focus the beam. This can be compensated with the focusing effect of the solenoid, but for example, it is not possible to focus on sample while decelerating below about 2\,keV with the magnetic field off (see fig.\,\ref{f6}, left). However, one can make use of cycloid-shaped modulations of the beam envelope in the case where the solenoid is powered. For example, the lowest magnetic field that can be used when soft-landing at 500\,eV is 1.70\,Tesla (see. fig.\,\ref{f6}, middle).
\begin{figure}[htbp]\centering
\includegraphics[height=4.5cm]{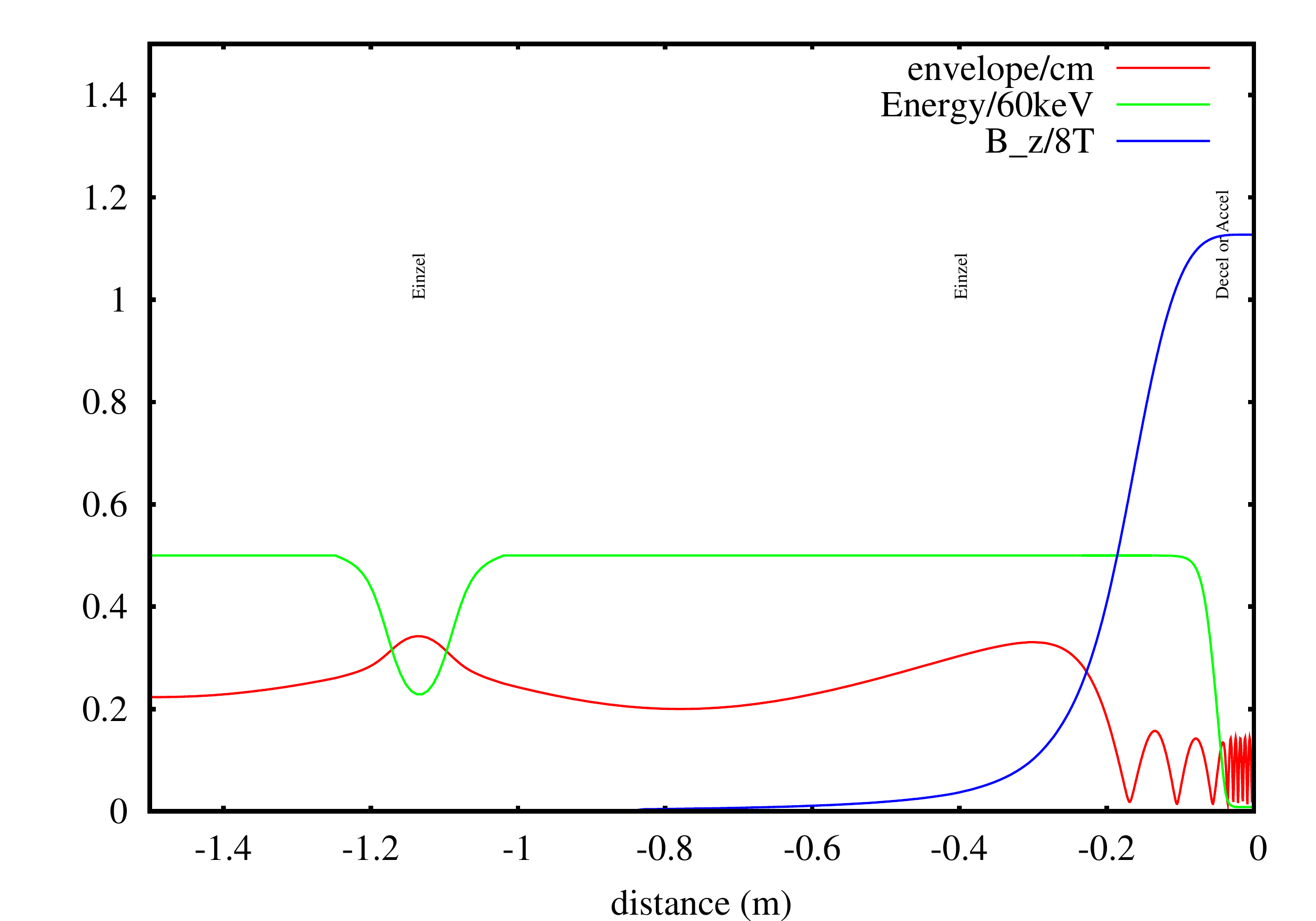}
\includegraphics[height=4.5cm]{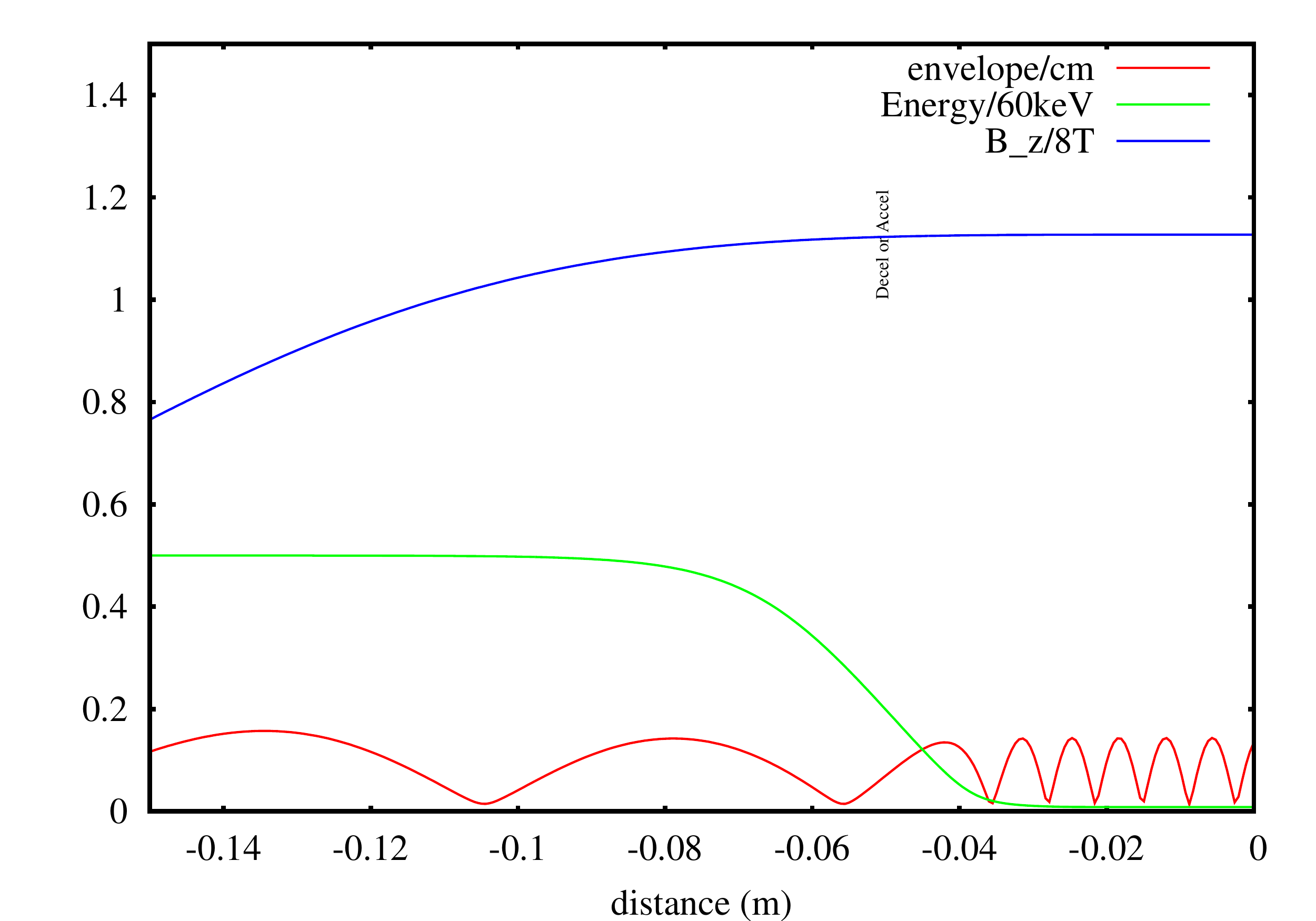}
\caption{Envelopes for emittance of $\pi\epsilon=12.5\,\pi\mu$m, for the $\beta$-NMR platform; the plot on the right is with expanded $z$ scale. The initial beam energy is 30\,keV, decelerating for soft landing on the sample to 500\,eV. The superconducting solenoid is at 9\,Tesla. The sample is mounted near $z=0$.}
\end{figure} 

\subsection{Polarizer}
The polarizer requires the radioactive ions to be neutralized and so they cannot be focused for a 2.5\,m drift length of beamline. This requires a matching section that expands the beam and decreases its divergence.
\begin{figure}[htbp]\centering
\includegraphics[height=3.2cm]{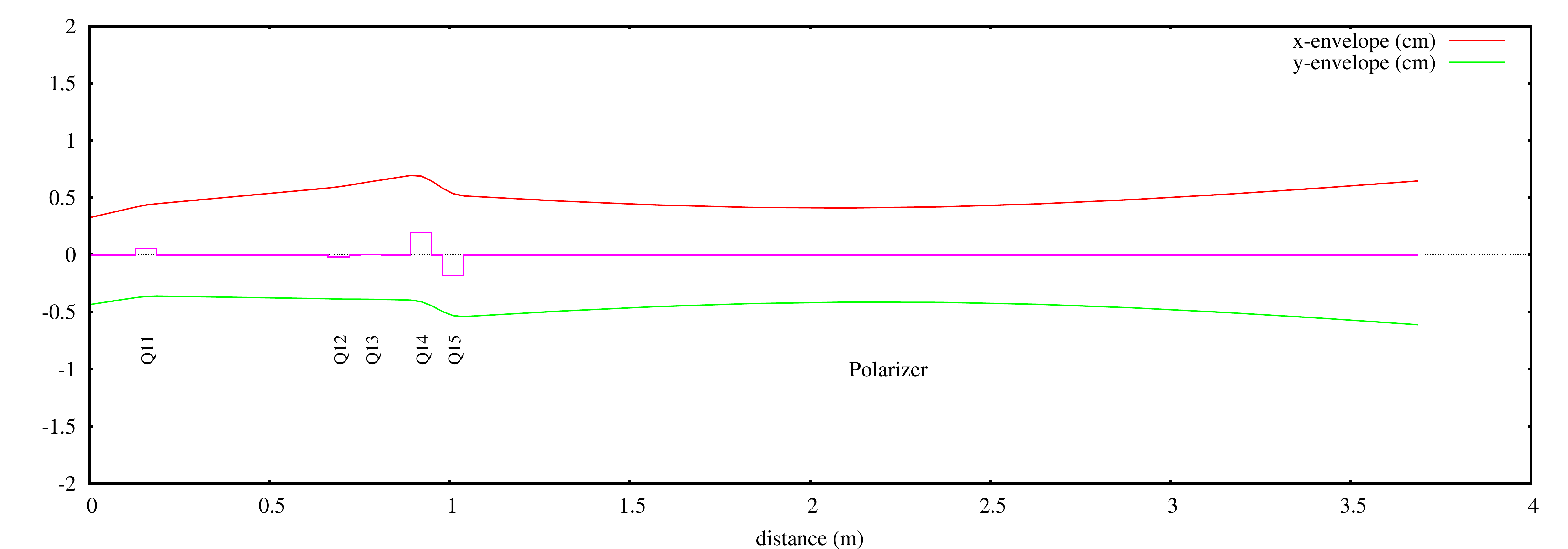}
\caption{Envelopes for section of beamline to match into a large, low divergence beam. $\epsilon=12.5\,\mu$m.}\label{f7}
\end{figure} 

\subsection{Charge State Booster}
\begin{figure}[htbp]\centering
\includegraphics[width=\textwidth]{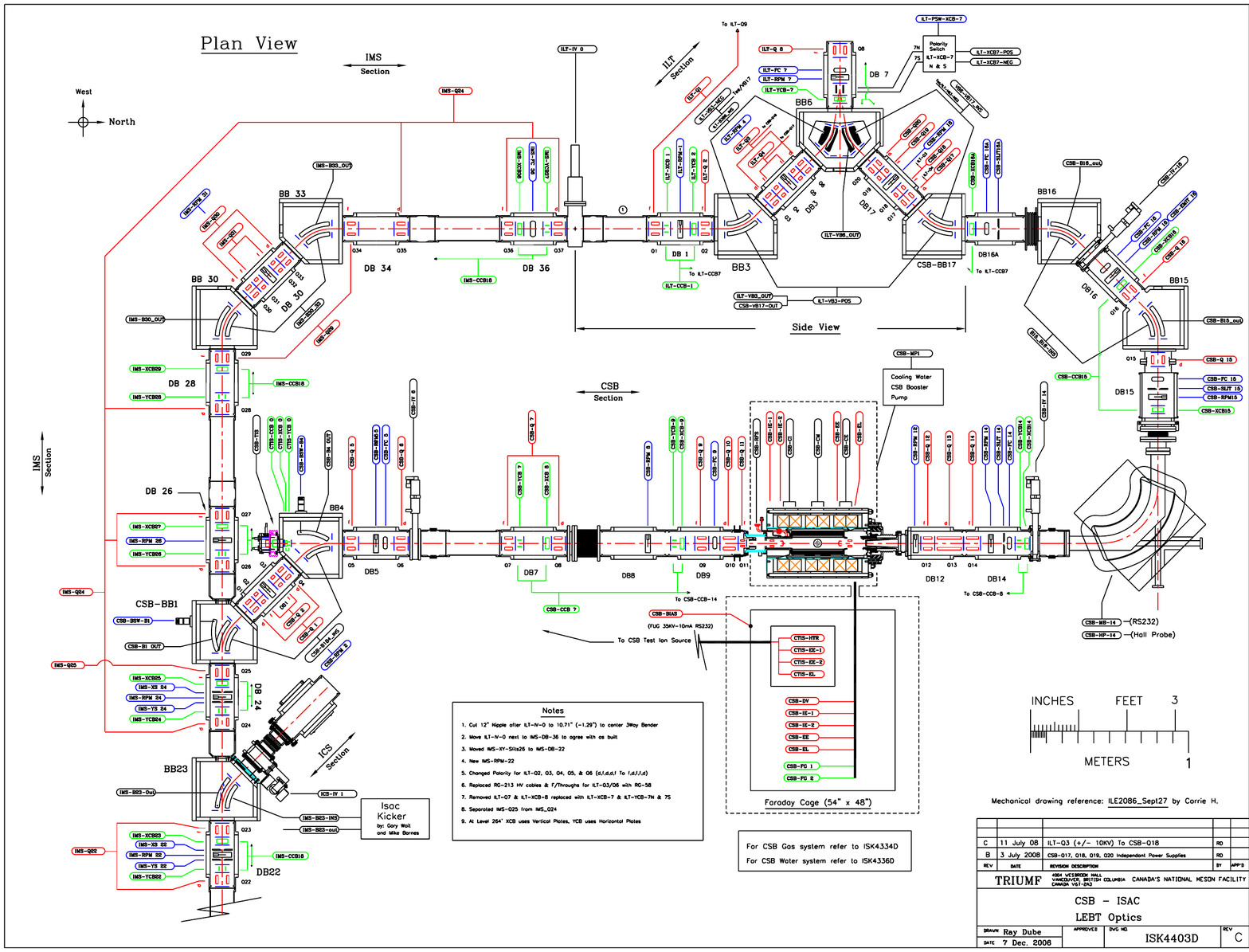}
\caption{Layout of the CSB. Note that most of the drawing is a plan view, but the bend system in the upper, right-of-centre portion is in a different plane, bending vertically.}\label{f8}
\end{figure}

The electron cyclotron resonance (ECR) charge state booster (CSB) was added in the separator area in 2007. See fig.\,\ref{f8}; it consists of a mechanical switch, matching to decelerate the beam to soft-land into the ECR, matching out, and a Nier-type separator and finally, the bender to the vertical section was modified to allow switch between the (old) bypass, or to send beam from the CSB.

\begin{figure}[htb]\centering
\includegraphics[width=.45\textwidth]{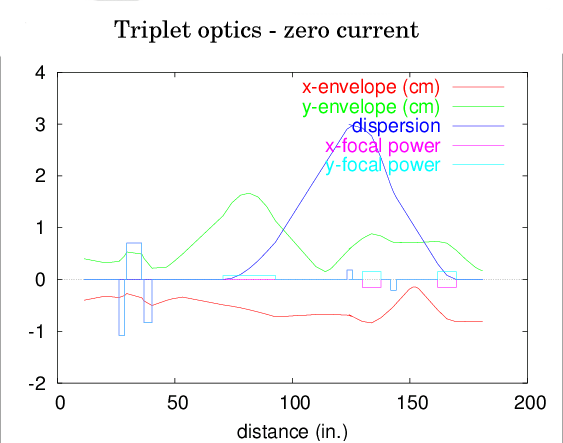}
\includegraphics[width=.45\textwidth]{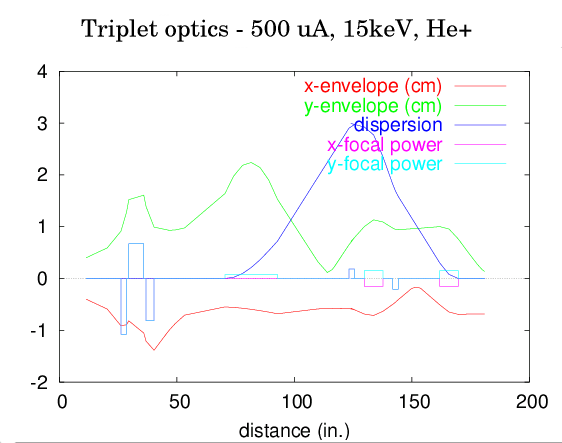}
\caption{Beam envelopes for two different levels of space charge, with equal settings of the quadrupoles. Left is for zero current, and right is at $500\,\mu$A.}\label{f9}
\end{figure}

Of note is the section of beamline from the ECR to the vertical section, which contains the Nier separator. From the ECR to the magnetic dipole, the ion beam is dominated by ions of the support gas, up to $500\,\mu$A, so space charge becomes an important effect. As this is the match into the Nier separator, a technique was needed to make the match independent of space charge. This was achieved by allowing the beam at the match point just upstream of the dipole to be large for high current cases, but small for low current cases. See fig.\,\ref{f9}.

\begin{figure}[htbp]\centering
\includegraphics[width=.95\textwidth]{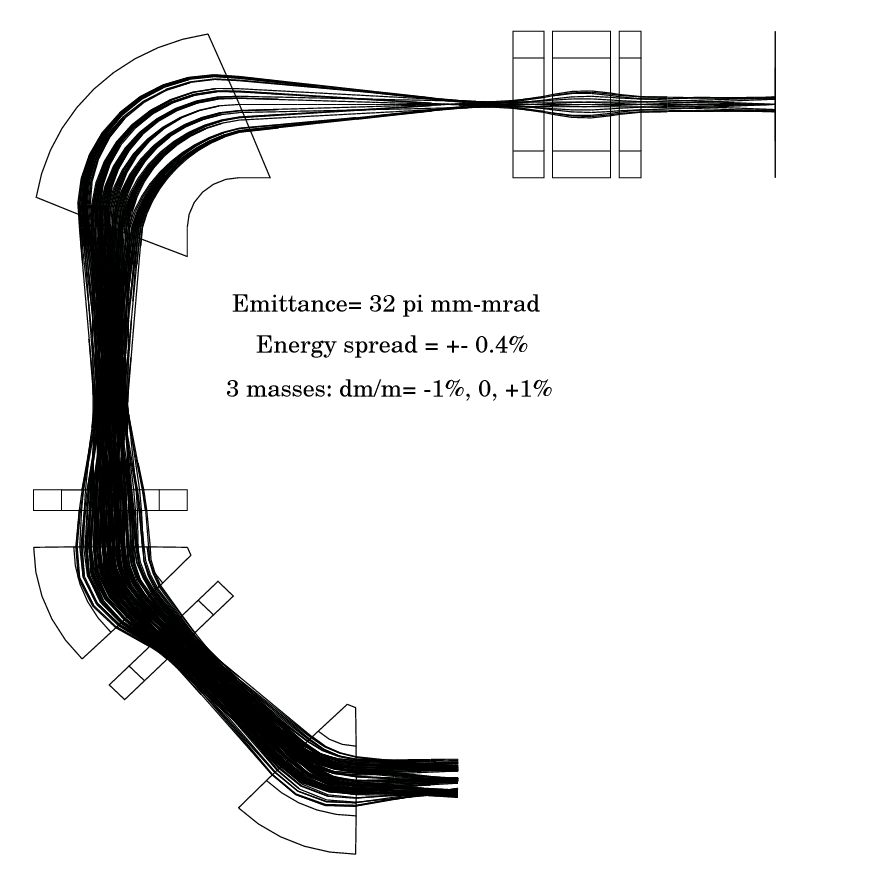}
\caption{Trajectories for three masses through Nier separator, as calculated to third order using {\tt GIOS}, showing a clear separation at resolution of 100 for a pessimistically large emittance and energy spread. (Note that the transverse size is not to scale, but enlarged for clarity. Beam proceeds from upper right, counterclockwise.)}\label{f10}
\end{figure}

The Nier separator has a total bend of $180^\circ$. The principle is that the energy dispersion of the magnetic bend is precisely compensated by the energy dispersion of the electrostatic bends. Since mass dispersions are different for the two types of bends, they do not cancel, resulting in a good mass separation in spite of a comparable fractional energy spread. Ours uses two standard ISAC LEBT electrostatic spherical bends and a magnetic dipole of largest possible size in the confined space. As mass resolution scales with size, the limited space allows a resolution of only roughly 200, depending upon ECR source emittance. See fig.\,\ref{f10}.
\end{document}